\documentclass[a4paper]{article}
\usepackage{amsmath,graphicx}
\usepackage{url}
\usepackage{amssymb}
\usepackage{rotating}

\title{Brain Correlates of Task-load and Dementia Elucidation with Tensor Machine Learning Using Oddball {BCI} Paradigm}
\author{Tomasz M. Rutkowski$^1$\thanks{Published as: 
T.~M. {Rutkowski}, M.~{Koculak}, M.~S. {Abe}, and M.~{Otake-Matsuura}, ``Brain
  correlates of task--load and dementia elucidation with tensor machine
  learning using oddball {BCI} paradigm,'' in {\em ICASSP 2019 - 2019 IEEE
  International Conference on Acoustics, Speech and Signal Processing
  (ICASSP)}, pp.~8578--8582, May 2019. DOI 10.1109/ICASSP.2019.8682387}, Marcin Koculak$^2$, \\Masato S. Abe$^1$, and Mihoko Otake-Matsuura$^1$
\\ \small $^1$Cognitive Behavioral Assistive Technology Team, RIKEN AIP, Tokyo, Japan\\
 \small $^2$Consciousness Lab, Institute of Psychology, Jagiellonian University, Krakow, Poland\\
 \small \url{tomasz.rutkowski@riken.jp}\\
 \small \url{http://aip.riken.jp/}}
 
 \date{}
\begin{document}
%
\maketitle
\begin{abstract}
Dementia in the elderly has recently become the most usual cause of cognitive decline. The proliferation of dementia cases in aging societies creates a remarkable economic as well as medical problems in many communities worldwide. A recently published report by The World Health Organization (WHO) estimates that about 47 million people are suffering from dementia-related neurocognitive declines worldwide. The number of dementia cases is predicted by 2050 to triple, which requires the creation of an AI-based technology application to support interventions with early screening for subsequent mental wellbeing checking as well as preservation with {\it digital--pharma} (the so-called {\it beyond a pill}) therapeutical approaches. 

We present an attempt and exploratory results of brain signal (EEG) classification to establish digital biomarkers for dementia stage elucidation. We discuss a comparison of various machine learning approaches for automatic event-related potentials (ERPs) classification of a high and low task--load sound stimulus recognition. These ERPs are similar to those in dementia. The proposed winning method using tensor-based machine learning in a deep fully connected neural network setting is a step forward to develop AI-based approaches for a subsequent application for subjective-- and mild--cognitive impairment (SCI and MCI) diagnostics. 
\end{abstract}
%
%

\section{Introduction}
\label{sec:intro}
Currently, dementia is one of the most significant global challenges for healthcare and social services. 
Worldwide, mainly for above 65 years old people, dementia statistics and associated costs are increasing due to enhanced longevity~\cite{lancet2017dementia}. Cabinet Office in Japan, to address the problem,
publishes annual reports on aging society~\cite{agingJPgovREPORT}. United Nations Sustainable Development Goal number three -- ``Good Health and Wellbeing" focus, for all at all ages, on healthy lives and it supports well-being.
We propose a method utilizing AI--related machine learning (ML) for automatic classification of anomalous EEG brain signals, which shall lead to digital biomarkers development for the task--load as well as dementia progress elucidation. The current dementia diagnostic methods rely on standard psychometric subjective tests, more recent behavioral assessments using cognitive behavioral therapy (CBT) methodologies  (a co--imagination technique~\cite{otake2009coimagination}, etc.) or the very recent multimodal interventions~\cite{kivipelto2018world}. We propose to use a human noninvasive brain signal monitoring (EEG) technique (namely, event-related potentials - ERPs derived from responses to natural stimuli within modern, comfortable recording~\cite{bciBOOKwolpaw,nozomuANDtomekAPSIPA2012,tomekJNM2015,tomekFRONTIERS2016}. Methods presented in this paper and developed by our lab, establish a machine learning (ML) or AI application allowing for near future a home monitoring of the mental decline (dementia) and the task--load elucidation with employment of digital biomarkers. 

Recent neuroscience research findings have shown that dementia decline was related to abnormal amyloid and tau disposals that modified post-- and presynaptic neuronal mechanisms, which resulted in high neuronal calcium influxes leading to neuronal loss, increased excitability and finally modified brain rhythmic patterns~\cite{oscillopathy2006,dementiaBIOMARKER2018review}. The synaptic plasticity is necessary for various cognitive functions (e.g., memory formation, abstract thinking, learning,  etc.). 
Dementia and especially Alzheimer's disease (AD) has been recently identified from contemporary neurophysiological and pathological studies as impaired synaptic plasticity~\cite{dementiaBIOMARKER2018review}. 
The above mentioned age-related brain atrophy has also been named as network disconnection disease called ``an oscillopathy''~\cite{oscillopathy2006}.

\begin{figure}
	\centering
  	\includegraphics[width=0.7\linewidth]{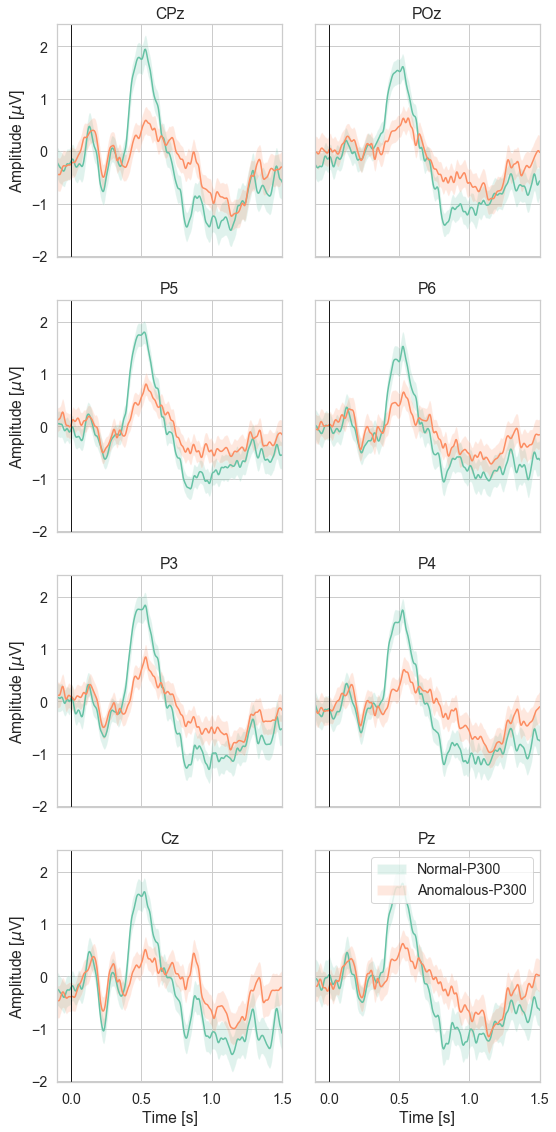}
	\caption{Grand mean averaged ERPs from nine healthy subjects performing spatial auditory BCI oddball experiments~\cite{nozomuANDtomekAPSIPA2012}. All EEG channels, as in classical $10/20$--system~\cite{bciBOOKwolpaw}, are presented as bandpass filtered and averaged traces with 95--percentile ranges. Green traces represent easy (low task--load) ERPs to identified spatial targets with clear P300 responses~\cite{bciBOOKwolpaw,nozomuANDtomekAPSIPA2012}. The orange ERPs to the difficult virtual cases (high task--load), which model dementia decline signals~\cite{dementiaBIOMARKER2018review}.}\label{fig:p300}    
\end{figure}

Recently,  various approaches to AD and dementia diagnostics created a necessity to establish personalized therapies using not only traditional pharmacological interventions but also the necessary lifestyle alternations~\cite{lancet2017dementia} as well as the very popular cognitive training~\cite{otake2009coimagination}. 
The usual pharmacological as well as the modern {\it digital--pharma} ({\it beyond a pill}) healthcare approaches call for robust biomarkers. The biomarkers need not only to elucidate the cognitive loss from resting-state-- digital brain biomarkers~\cite{tomekICASSP2006} but must allow for simple home monitoring. We evaluate the so-called work-- task--load paradigm for dementia-related brain responses' classification in natural auditory stimulus settings. In this paper we report on the evaluation of EEG  classification methods using dementia--modeling ERPs~\cite{oscillopathy2006,dementiaBIOMARKER2018review} in spatial auditory and task--load controlled brain-computer interface (BCI) experimental settings and from healthy subjects~\cite{nozomuANDtomekAPSIPA2012}. 
The spatial auditory BCI paradigm developed in~\cite{nozomuANDtomekAPSIPA2012} allows for the task--load modulation by incorporating real and virtual spatial source sounds, which are easy or hard to localize spatially. 
The users are instructed to focus their attention on spatial targets (resulting with P300 responses) and ignore distractors (a classical oddball paradigm)~\cite{nozomuANDtomekAPSIPA2012,tomekJNM2015,tomekFRONTIERS2016}. 
Averaged results with abnormal and normal (varying task--load settings to model dementia responses~\cite{oscillopathy2006,dementiaBIOMARKER2018review}) ERPs are depicted in Figure~\ref{fig:p300}.

We propose a framework, using the AI--based neurotechnology, which employs ML together with data--driven preprocessing techniques utilizing a wavelet--synchro--squeezing--transform (WSST)~\cite{tomekJNM2015,tomekJOMS2016}, Riemmanian geometry (RG)~\cite{barachant2010riemannian} and tensor--train--layer (TT--layer)~\cite{oseledets2011tensor,cichocki2017tensor} classification methods.

From now on the paper is organized as follows. In the next section, we introduce EEG signal processing and machine learning methods. Results and conclusions with future research guidelines summarize the paper.

\section{Methods}\label{sec:methods}
The brain responses, which model EEG--derived digital biomarkers in our study, were collected from nine healthy subjects in a spatial auditory BCI former project~\cite{nozomuANDtomekAPSIPA2012} reviewed by The Ethical Committee of Faculty of Engineering, Information and Systems at the University of Tsukuba, Tsukuba, Japan.  The data collection details were described in~\cite{nozomuANDtomekAPSIPA2012}.
The currently discussed EEG brain signal post-processing project was approved by The Ethical Committee of RIKEN, Wako--shi, Japan. 

At the EEG preprocessing filtering steps we utilize a wavelet--synchro--squeezing--transform (WSST) method previously published by the authors for sleep stages' and BCI commands'  classification~\cite{tomekJOMS2016,tomekJNM2015}. The WSST application follows precisely the cased discussed in~\cite{tomekJNM2015}.
At the next step of EEG processing pipeline we assume that $\mathbf{x}(t)\in\mathbb{R}^N$ is a zero--mean data sample captured from eight (denoted by $N$) EEG electrodes at discrete time $t.$ 
In such a case, let $\mathbf{x}_{c,i}\in\mathbb{\mathbb{}R}^{N}$ be the ERP event $i$ representing a response to a spatial auditory stimulus $c\in \{1,2\}$ of $M$ samples ($1.5$~seconds in our case). 
Assuming a a zero mean, a sample covariance matrix of a single trial $\mathbf{x}_{c,i}$ belonging to a class $c$ is given by $\mathbf{C}_{c,i} = \frac{1}{M-1}\mathbf{x}_{c,i} \mathbf{x}_{c,i}^T,$ as proposed by~\cite{barachant2010riemannian}.

Assuming that a noise in EEG recordings could be modeled by multivariate Gaussian distributions, a covariance matrix characterizing ERP features could be considered as the only unique parameter for dementia stages (task--load). Features representing stimulus in segmented ERPs $c\in \{1,2\}$ are obtained as $\mathbf{x}_{c,i}$ and calculated as covariance matrices $\mathbf{C}_{c,i}$ for a subsequent input to ML, as discussed in following sections. 

\subsection{Shallow Learning Classifiers}
\label{sec:shallow}

In a classifier training phase~\cite{barachant2010riemannian}, a geometric mean covariance matrix $\mathbf{\overline{C}}_{c},$ characterizing each ERP class $c,$ is computed, using all EEG channels as inputs. In order to measure a distance of a newly captured ERP to the class--characterizing mean matrix, the RG techniques are used.  
A geodesic traversing two points $\mathbf{C}_i$ and $\mathbf{C}_j$ is the shortest path curve that connects them. It has a minimum length. A Riemannian distance between covariance matrices is obtained as follows, 
\begin{equation}
    \delta_R = \left|\left|ln(\mathbf{C}_i^{-1}\mathbf{C}_j)\right|\right|_F = \sqrt{\sum_n[ln(w_n)]^2},
\end{equation}
where $||\cdot||_F$ denotes a Frobenius norm and $w_1,\ldots,w_n$ the eigenvalues of $\mathbf{C}_i^{-1}\mathbf{C}_j$, respectively~\cite{barachant2010riemannian}. 

The geometric mean of $L$ covariance matrices characterizing an ERP class is obtained as, 
\begin{equation}
    D(\mathbf{C}_1,\cdots,\mathbf{C}_l) = \arg\min_{\mathbf{C}}\sum_{l=1}^L\delta_R^2(\mathbf{C},\mathbf{C}_l). 
\end{equation}
On a manifold, the geodesic, according to the RG principles~\cite{barachant2010riemannian} is calculated from,  
\begin{equation}
    \Gamma(\mathbf{C}_i,\mathbf{C}_j, \tau) = \mathbf{C}_i^{\frac{1}{2}}\left(\mathbf{C}_i^{-\frac{1}{2}}\mathbf{C}_j\mathbf{C}_i^{-\frac{1}{2}}\right)^\tau\mathbf{C}_i^{\frac{1}{2}},
\end{equation}
where  $\tau\in\{0,1\}$ is a scalar.

\begin{sidewaysfigure}
	\centering
  	\includegraphics[width=\linewidth]{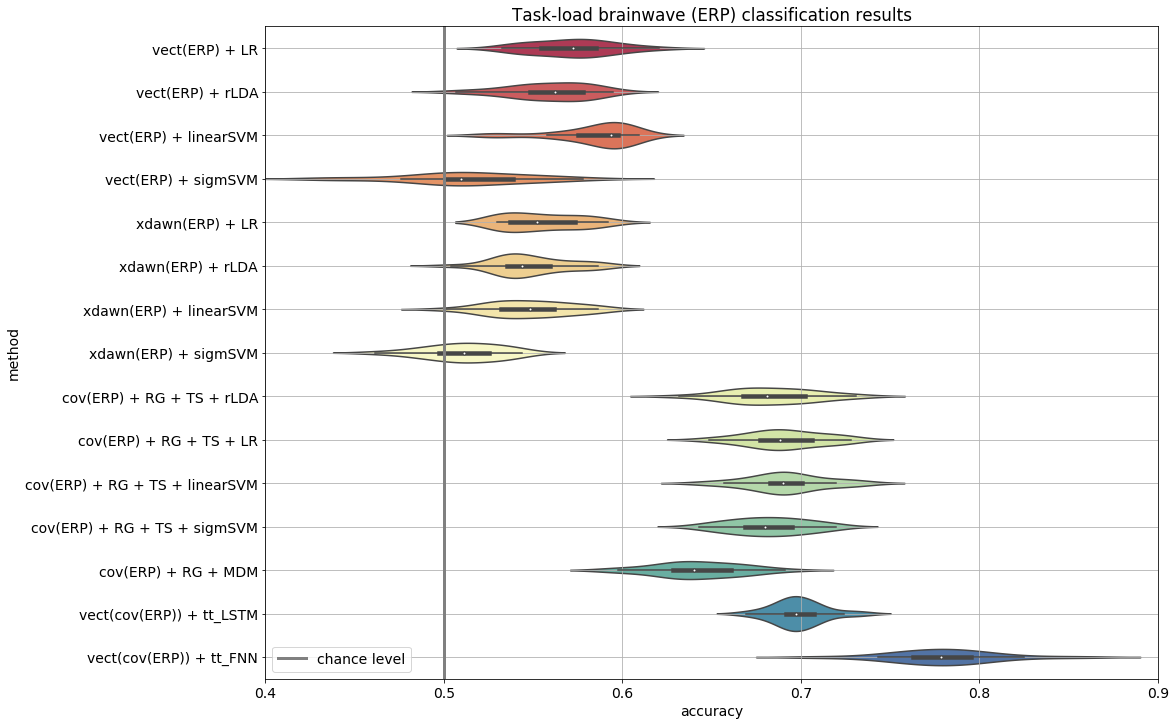}
	\caption{Median classification accuracy results (white dots within percentile ranges) depicted together with result distributions (a chance level of $0.5$ representing $50\%$ of binary classification cases depicted with bold gray line). }
	\label{fig:acc}    
\end{sidewaysfigure}

A very frequently utilized classification approach for RG--related features has been based on a distance evaluation among ERP features and mean (class characterizing) covariance matrices~\cite{barachant2010riemannian}. A a minimum distance to mean (MDM) classifier~\cite{barachant2010riemannian} satisfies the above criterium. 
The MDM approach is easy to apply as well as very generic. We compare it with previously reported by our group in~\cite{tomekNIPS2018} classical vectorized (vect) time domain derived ERP features~\cite{tomekJNM2015}; spatial filter xDAWN preprocessing~\cite{hiroshiAPSIPA2016}; tangent space (TS) mapping of RG features~\cite{barachant2010riemannian}; using linear regression (LR); regularized linear discriminant analysis (rLDA); linear and sigmoid kernel support vector machine (SVM), as well as with below to be introduced TT-layer--based (TT) methods as shown in Figure~\ref{fig:acc}.

\subsection{Tensor--train--layer--based Neural Network Classifiers}
\label{sec:tt}

Deep neural networks recently started to gain attention in EEG community especially for the end--to--end processing setups~\cite{braindecode2017}. For our approach, we are interested in compacting machine learning architectures with possible application to wearable devices with limited computing powers.
Tensor--train factorization (TTF) approaches have demonstrated an advantage of scaling multidimensional matrices to an arbitrary number of dimensions~\cite{oseledets2011tensor,cichocki2017tensor} with a subsequent possibility to reshape a fully connected neural network layer into a tensor and then factorize it~\cite{novikov2015tensorizing}. The methodology as mentioned earlier is applied to compress large weight matrices of deep neural models within an entire end--to--end training. Using a trick from~\cite{novikov2015tensorizing} it is possible to tensor--train--factorize a weight matrix $\mathbf{W}$ of a fully connected neural network as a $d-$dimensional double--indexed tensor represented as
\begin{eqnarray}
    \widehat{\mathbf{W}}\left((i_1, j_1), (i_2, j_2),\ldots,(i_d, j_d)\right) = \nonumber \\ 
    = \mathbf{\mathcal{G}}^*_1(i_1, j_1)\mathbf{\mathcal{G}}^*_2(i_2, j_2) \ldots\mathbf{\mathcal{G}}^*_d(i_d, j_d), 
\end{eqnarray}
where $\mathbf{\mathcal{G}}^*_k \in \mathbb{R}^{m_k \times n_k \times r_{k-1} \times r_k}$ are the so--called core tensors uniquely represented  by indices $(i_k, j_k)$~\cite{novikov2015tensorizing}. It is possible to compress the weight matrix $\mathbf{W}$ of size $\prod_{k=1}^d m_k  \cdot n_k$ in a form of TT--formated low--rank tensors, approximately reconstructing the original $\mathbf{W}$, of size $\sum_{k=1}^d m_k \cdot n_k  \cdot r_{k-1} \cdot r_k.$
We consider a two--layered neural network with $64$ hidden units in each layer and replace both fully-connected layers by the TT--layers with all the TT--ranks in the network set to $4$ and activation functions to ReLU. For an output of a fully connected final layer with sotfmax activations we choose two units. In order to prevent overfitting we train and test (using ``unseen'' during the training samples) the model with $50\%/50\%$ balanced datasets.
We also evaluate a recurrent neural network with LSTM units as a classifier with TT--layers applied to gating units~\cite{yang2017tensor}. In this case the network has again $64$ input LSTM units and TT--ranks set also to $4$. A fully connected output layer with sotfmax activations also has two units in this case. 

\section{Results}
\label{sec:results}

The discussed neurotechnology approach utilizing  AI/ML technique for the task--load elucidation, which has been proposed as a model for demented brain cognitive responses quantification, resulted in a particular classification increase. 
We compared the information geometry principle using shallow learning and TT--layer classification methods introduced in the previous section. 
A  solid, as well as statistically significant, the boost of the automatic task--load--dependent ERP classification was achieved in the presented project. The project goal has been to automatically discriminate the EEG signals modulated by two tasks--load--dependent cognitive levels, comprising of ordinary (easy) and anomalous (difficult) spatial auditory localization tasks.
We obtained results from the nine subjects--based dataset with classifiers trained using all participant class--balanced, to avoid overfitting, training sets.
Subsequently, we tested classifiers with respective testing utterances.
The results were summarized in the form of classification accuracy distribution plots in Figure~\ref{fig:acc}.

The best results among the shallow learning classifiers (see Section~\ref{sec:shallow}) with an average of $68\%,$  which has been considered already as satisfactory for BCI applications, were obtained for the tangent space (TC) mapping with the linear regression (LR), the regularized LDA (rLDA) and support vector machine (SVM) cases, comparing to the minimum distance to mean (MDM) and to the simple vectorized time domain ERP features. The differences were statistically significant in all comparisons as evaluated with non--parametric pairwise Wilcoxon method. RG approaches scored significantly better comparing to the classical vectorized (vect) and xDAWN--based results with $p \ll 0.01.$ Also the RG tangent space (TS) mapping--based results were significantly better, with  $p < 0.05$,  than RG--based MDM classifier accuracy results, respectively. The shallow learning results confirmed our previous findings reported in~\cite{tomekNIPS2018}.

On the other hand, using the TT--layer--based LSTM (see Section~\ref{sec:tt}), we have achieved median accuracy of $68\%$ for two task--load cases with significant differences $(p\ll 0.01)$ compared to all other shallow learning classifiers (see Figure~\ref{fig:acc} with accuracy comparisons), as evaluated with non--parametric Wilcoxon pairwise tests. 

Finally, the results achieved with TT-layer fully connected neural network (details also in Section~\ref{sec:tt}) had a median accuracy of $78\%$ for two task--load cases with significant differences $(p\ll 0.01)$ compared to all other tested in this paper classifiers (see Figure~\ref{fig:acc}), as also calculated using Wilcoxon pairwise comparisons.

\section{Conclusions}
\label{sec:conclusions}

The presented project confirmed our hypothesis of a possible application of  TT--layer neural model--based task--load elucidation approach as tested with nine subjects. The proposed approach resulted with a satisfactory classification ($78\%$ on average) of binary task--load cases, namely the 'easy real' versus the 'difficult virtual' sound localization paradigms, as shown as averaged ERPs in Figure~\ref{fig:p300}, and the obtained classification accuracy boost results in Figure~\ref{fig:acc}. 

The discussed EEG--task--load classification results offer a step forward in the establishment of novel dementia-related biomarkers for improving the lives of seniors and reducing health-related costs. The task--load responses (EPRs) were very similar to EEG modulations in dementia. 

As for following steps in our project we plan to test the discussed method with elderly healthy and mild cognitive impairment (MCI), dementia, etc., subjects. 
We also plan to collect more data to further train and evaluate deep learning models in order to achieve improvements in the proposed approach towards the final application in the real--world healthcare settings. Further research on the presented TT--layer neural model interpretability and a practical computational cost are also planned.


\end{document}